\documentclass[9pt,twocolumn,twoside]{osajnl}

\journal{josab} 

\setboolean{shortarticle}{false} 

\ifthenelse{\boolean{shortarticle}}{\colorlet{color2}{color2b}}{\colorlet{color2}{color2}} 
\title{Suppression of Four-Wave Mixing in Hot Rubidium Vapor Using Ladder Scheme Raman Absorption}
\author{Nikunj Prajapati*}
\author{Gleb Romanov}
\author{Irina Novikova}

\affil{Department of Physics, College of William $\&$ Mary, Williamsburg, Virginia 23187, USA}

\affil[*]{Corresponding author: nprajapati@email.wm.edu}

\dates{Compiled \today}

\ociscodes{(020.1670) Coherent optical effects; (020.4180) Multiphoton processes; (270.1670) Coherent optical effects.}

\doi{\url{http://dx.doi.org/10.1364/ao.XX.XXXXXX}}

\begin{abstract}
We experimentally investigate the effectiveness of the four-wave mixing suppression in a double-$\Lambda$ interaction scheme by introducing an additional ladder-type two-photon Raman absorption resonance for one of the optical fields. We propose several possible interaction configurations involving either one or two isotopes of Rb, and experimentally demonstrate the possibility of efficient four-wave mixing suppression in both EIT and far-detuned Raman cases. 
\end{abstract}

\setboolean{displaycopyright}{true}

\begin{document}

\maketitle
\thispagestyle{fancy}

\ifthenelse{\boolean{shortarticle}}{\ifthenelse{\boolean{singlecolumn}}{\abscontentformatted}{\abscontent}}{}

\section{Introduction}

Strong coupling of light with long-lived collective spin states of atomic ensembles~\cite{lukin03rmp} is a key component of many quantum information protocols. It enables light-matter entanglement that is the heart of some quantum repeater protocols~\cite{DLCZ,GizinRepeatersRMP11} and is a required condition for the realization of quantum memory~\cite{lvovskyNPh09,euromemory,quantmemoryreviewJMO2016}. While many memory protocols have been proposed and successfully demonstrated in the recent years, many of them rely on the reversible mapping of an optical quantum state into the atomic coherence between hyperfine states of alkali metal atoms via a two-photon Raman transition. This allows significant extension of storage time; all the while, a strong control optical field often gives rise to four-wave mixing (FWM), a process in which the stored optical signal is incoherently amplified and a new stokes optical field is generated~\cite{lukin97prl,narducciPRA04,HaradaPRA08,EilamOL08,hong09,Howell_2009,phillipsPRA11,gengNJP2014}. The additional quantum noise associated with the four-wave mixing process has been shown to severely limit the fidelity of several quantum memory systems~\cite{laukPRA13,DabrowskiOE14,MichelbergerNJP15}. Several approaches have been proposed to reduce the negative effects of the four-wave mixing by optimizing  frequencies ~\cite{HaradaPRA08,bashkanskyPRA13} or polarizations~\cite{zhangPRA14}, or introducing an optical cavity for spectral filtering~\cite{nunnRamancavityPRL16,nunnRamancavitytheory16arxiv}, but each comes with its limitations.

Recently, we have proposed to use resonant Raman absorption for the newly generated optical field to limit the overall FWM gain~\cite{romanovJMO2016}. Theory predicted that with sufficient stokes absorption, it should be possible to effectively suppress the optical probe amplification from the four-photon process (FWM) without affecting the two-photon interaction via electromagnetically induced transparency (EIT)~\cite{harris'97pt,marangos'98,lukinRMP03,fleishhauerLukinPRL00,fleishhauerLukinPRA02,novikovaLPR12}. However, the experimental realization of EIT-based stored light and narrowband absorption resonances for the stokes field using two Rb isotopes introduced additional technical difficulties, such as the need for isotope mixture optimization or the need to suppress the unwanted resonant effects of Raman pump field on the atoms of the ``storage'' isotope. In this paper, we extend our previous proposal to rely on Raman absorption in a two-photon ladder configuration to suppress the generated stokes field which, in principle, helps alleviate many challenges of our previous scheme. Since the wavelength of the optical field required in this case to create stokes absorption resonance differed significantly from that of the optical fields involved in memory-related transitions, it did not directly affect the ground-state coherence. It is also maybe possible, at least in principle, to use the same atomic isotope for all transitions.

\section{Experimental arrangements}

In this manuscript we tested two interaction configurations widely used in quantum memory experiments. In the first case a strong control field and a weak probe field form a resonant $\Lambda$ system, the configuration commonly used to realize the EIT quantum memory~\cite{fleishhauerLukinPRL00,fleishhauerLukinPRA02,novikovaLPR12}. In the second case, two optical fields are far-detuned from any optical resonances, while remaining in a two-photon resonance. This arrangement closely resembles the interaction scheme used for the off-resonant Raman memory experiments~\cite{ReimPRL11}. In both cases the additional scattering of the strong control field off the ground-state coherence at the probe field's optical transition results in the generation of a new stokes optical field in a double-$\Lambda$ four-photon resonance. As it was shown before, both theoretically and experimentally, this additional FWM interaction results in the incoherent amplification of the original probe field, leading to uncorrelated excess quantum noise in the quantum memory channel. An in situ resonant absorption for the newly-generated stokes field suppresses the four-wave mixing. In both configurations we rely on the Raman transition to the second excited electronic state, enabled by an additional Raman pump optical field in a ladder configuration, to create a strong absorption exclusively for the stokes field. Simultaneously, we should pay particular attention so that this additional laser field does not modify optical propagation of either control or probe fields, to avoid its effect on the potential quantum memory performance.

\begin{figure}[H]
	\centering
		\includegraphics[width=1\columnwidth]{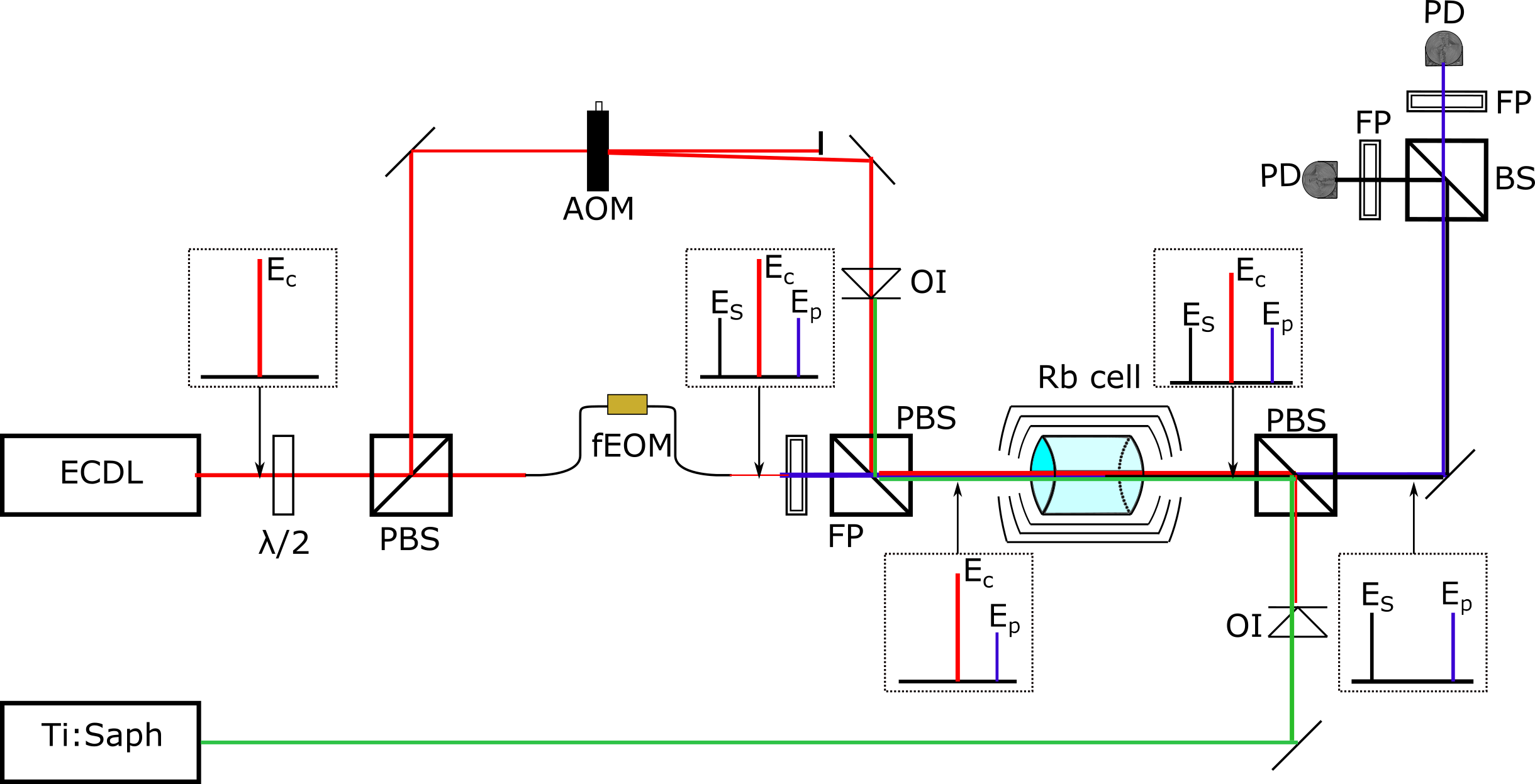}
	\caption{Optical schematic of setup. ECDL and Ti:Sapph denote the two independent lasers used in the experiment (external cavity diode laser and Ti:Sapph cw laser, correspondingly). Optical path of the control field is shown in red, fpr the Raman pump laser - in green, and the probe and stokes fields are correspondingly blue and black. See text below for abbreviations. }
	\label{fig:setup}
\end{figure}

Since both interaction schemes are quite similar, we can use the same basic experimental setup to test both of them. The schematic of the experimental setup is shown in Fig.~\ref{fig:setup}. 

Since the relative phase coherence between the control and probe fields is crucial for the quality of two-photon resonances, we derive both of these fields from a single laser (external cavity diode laser, or ECDL) tuned to the $D_1$ line of Rb (wavelength $794.6$~nm). The probe field is produced by phase-modulating a fraction of the laser output by a fiber electro-optical modulator (fEOM) and filtering out one of the first-order modulation sidebands using a tunable fabri-perot etalon (FP) with $20$~GHz free spectral range. The remaining laser output passed through an acousto-optical modulator (AOM), and the $+1$ modulation sideband was used as a control field. The control and the probe field were recombined at a polarizing beam splitter (PBS) before interaction with atoms. Maximum available control power was $\approx 30$~mW, and the power of the probe field was $140~\mu$W. After the cell, the control field was filtered out by another PBS, and the remaining optical beam was sent to a non-polarizing beam splitter (BS), two outputs of which were directed into two independent fabri-perot etalons, tuned to transmit correspondingly the probe and stokes fields. 

A cw Ti:Sapphire laser (Ti:Sapph) tuned to the $5P \rightarrow 5D_{3/2}$ transition of Rb (wavelength $762.1$~nm) was used as a Raman pump field. It was combined with the rest of the optical fields at the second polarizing beam splitter and traversed the cell in a counter-propagating direction to minimize the Doppler broadening of a two-photon resonance. All laser beams were weakly collimated inside the cell to the diameter of $1$~mm. Since all the optical fields were nearly collinear inside the cell, two optical isolators (OI) were placed to protect both lasers from the incoming strong pump beams. 

For the experiments, described below, we used a Pirex cylindrical cell (diameter $25$~mm, length $75$~mm)  containing natural abundance Rb isotope mixture. It was placed inside a three-layer magnetic shielding to suppress stray magnetic fields. The temperature of the cell was actively stabilized at $90^\circ$C using an electrical heater wrapped around the innermost layer of the magnetic shielding. The  corresponding atomic densities were $1.7\cdot{10}^{12}~\mathrm{cm}^{-3}$ for ${}^{85}$Rb and $0.7\cdot{10}^{12}~\mathrm{cm}^{-3}$ for ${}^{87}$Rb. 

\section{Resonant EIT case}

\begin{figure}[h]
	\begin{center}
		\includegraphics[width=1.0\columnwidth]{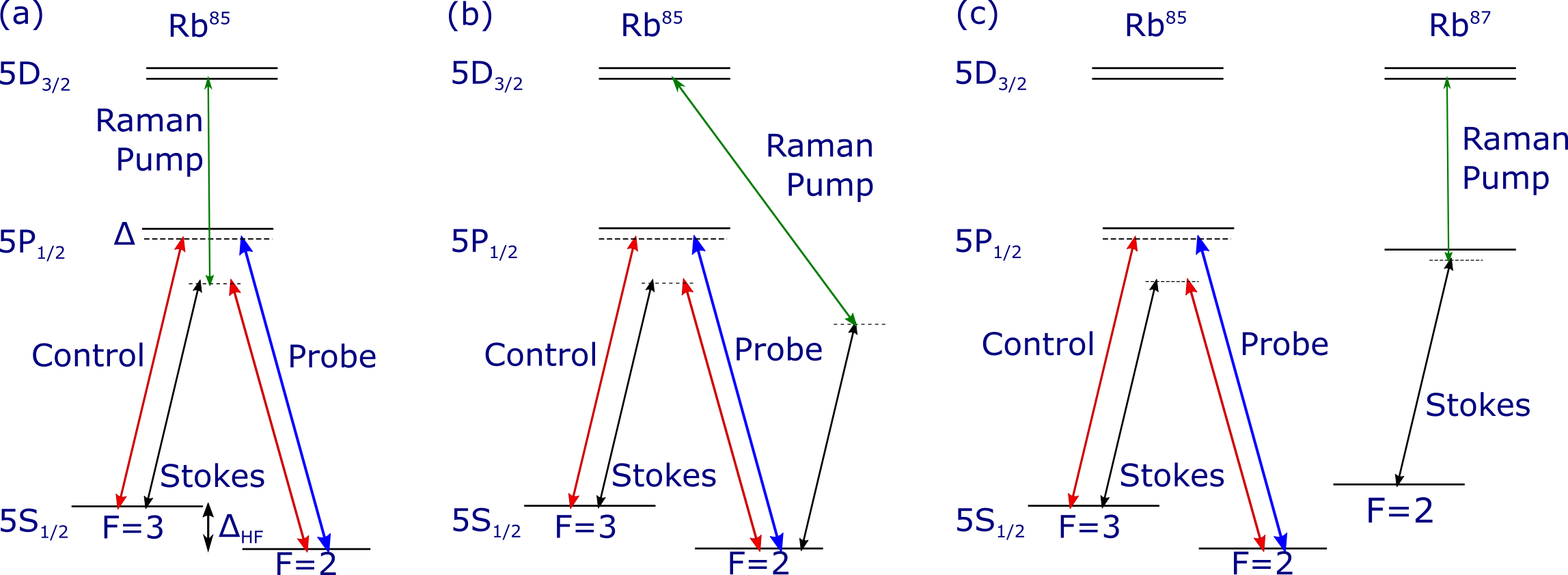}
		\caption{Possible realizations of a ladder Raman absorption resonance for the stokes field in EIT configuration using only ${}^{85}$Rb atoms \emph{(a,b)}, or  using ${}^{85}$Rb for EIT and ${}^{87}$Rb for Raman absorption \emph{(c)}. $\Delta$ is $80$~MHz, and $\Delta_\mathrm{HF}$ is 3035MHz.}
		\label{fig:eit_levels}
	\end{center}
\end{figure}

EIT configuration corresponds to both control and probe fields' frequencies tuned near optical resonances. For pure EIT we would expect to observe an increase in the probe's transmission when the two-photon detuning matched the hyperfine splitting between two Rb ground states~\cite{lukin03rmp}. The width of this resonance, as well as the residual absorption was determined by the strength of the control field and the decoherence rate of the ground-state coherence. The co-existing four-wave mixing typically increases the height of the probe field transmission due to additional gain. Simultaneously, it enables generation of an additional stokes optical field at the optical frequency shifted down by the hyperfine splitting from the control field, as shown in Fig. ~\ref{fig:eit_levels}. 
If the residual probe absorption under the pure EIT conditions is negligeable, under the combined EIT and FWM effects the probe output amplitude at the peak may exceed its initial value. Successfuls FWM suppression, in this case, should eliminate this additional gain; in the ideal case scenario, the output stokes field should completely disappear, while the probe transmission would diminish to the level determined only by the two-photon EIT resonance~\cite{romanovJMO2016}. Under the realistic conditions of limited control power, even at the EIT maximum the probe transmission is significant, and FWM gain does not elevate the signal level above its input value, so it is harder to distinguish as both processes add up coherently in the probe propagation~\cite{hong09,phillipsJMO09,PKLamNJP14}. However, the appearance of the stokes field in the same range of two-photon detunings when no input stokes field was present is a clear sign of the four-wave mixing process.   
The exact values of the FWM gain for both probe and stokes field depended on the mutual spatial alignment of the control and probe beams, and we normally adjusted the beams' positions to achieve higher power  and similar sensitivities to the control beam alignment for both probe and stokes outputs. 

For the EIT experiments the ECDL frequency was locked to the $5S_{1/2} F=3 \rightarrow 5P_{1/2} F'=3$ transition  of ${}^{85}$Rb using a separate reference cell (not shown in Fig.~\ref{fig:setup}), so thanks to the $+80$~MHz AOM-induced frequency shift the control field was tuned $80$~MHz above the $5S_{1/2},/F=3 \rightarrow 5P_{1/2} F'=3$ optical transition. To ensure that the frequency difference between the control and probe match the ${}^{85}$Rb hyperfine splitting $\Delta_(HF) = 3035$~MHz, the rf modulation frequency for fEOM was set on $\approx 3115$~MHz. By varying the frequency difference between the control and probe optical fields by sweeping the modulation frequency of the fEOM,  we observed clear transmission peak in the probe field around two-photon resonance conditions, as well as generation of the stokes field, marking the presence of the FWM effect.

\begin{figure}[H]
	\centering
		\includegraphics[width=1\columnwidth]{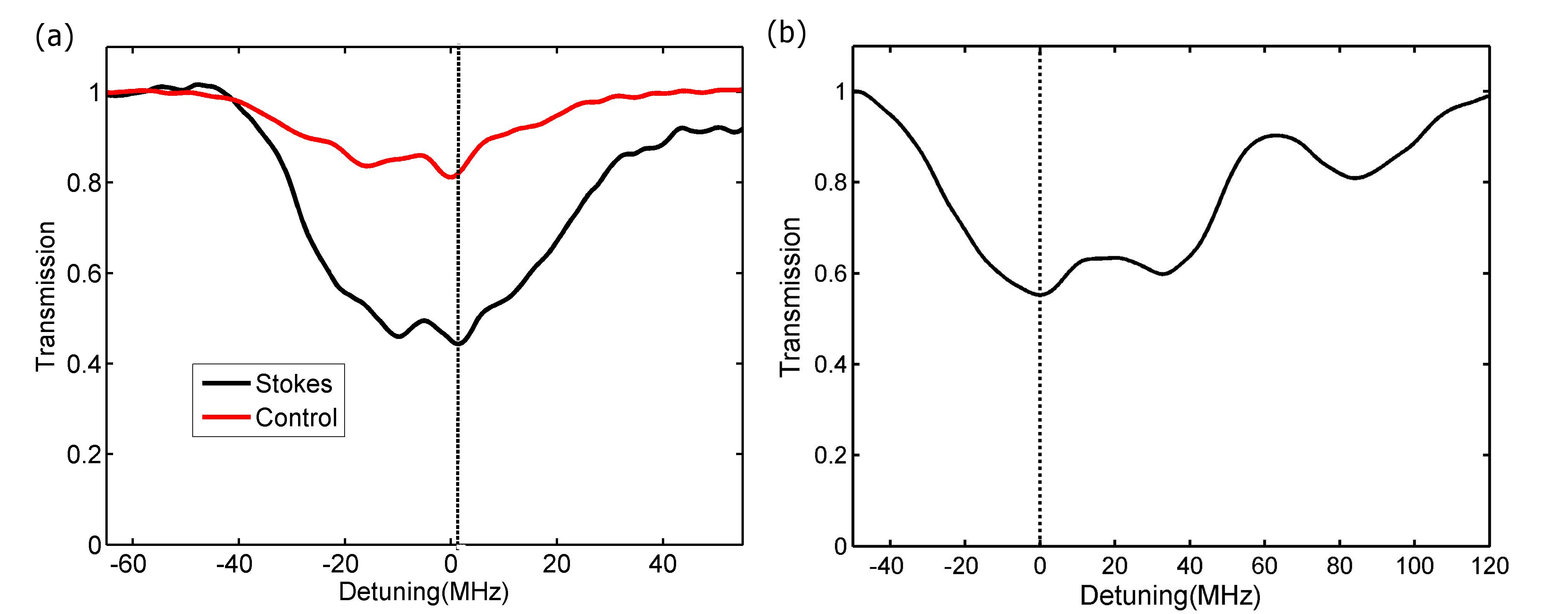}
	\caption{\emph{(a)} Measured transmission for stokes (black) and control(black) optical fields as the Raman pump frequency is scanned across the two-photon absorption resonance in the level configuration shown in Fig.~\ref{fig:eit_levels}(a). Both control and stokes fields experience absorption, since they are simultaneoulsy in two-photon resonance. \emph{(b)} stokes field absorption vs Raman pump frequency using ${}^{87}$Rb resonances, as shown in Fig.~\ref{fig:eit_levels}(c). Control field has no absorption in this case, and thus not shown. All curves are normalized to the transmission value without Raman pump. Vertical dashed lines indicate the optimal operational frequency. Raman pump power was $180$~mW for \emph{(a)} and $220$~mW for \emph{(b)}.} 
	\label{fig:control_abs_EIT}
\end{figure}

Possible realizations of the Raman absorption resonance for the stokes field in this configuration are shown in Fig.~\ref{fig:eit_levels}. If only one Rb isotope is involved, there are two possible arrangements. One is when the stokes field and the Raman pump field form a ``ladder'' from  $5S_{1/2},/F=3$ ground state to $5D_{3/2}$ second excited state, as shown in Fig.~\ref{fig:eit_levels}(a). In this case, the wavelength of the Raman pump field is $\lambda_\mathrm{pump} = 762.0976$~nm and produces the desired strong absorption resonance for the stokes field. Unfortunately, in this configuration the control field and the Raman pump field also form a ladder system, resulting in two-photon absorption of the control field. For instance, under the conditions when we observed $60\%$ stokes absorption, we also measured $20\%$ control field absorption, as shown in Fig.~\ref{fig:control_abs_EIT}(a). In principle, if sufficient control field power is available, such additional control absorption may not strongly affect the EIT interaction. However, a noticeable longitudinal variation of the control field power can lead to additional inhomogeneous broadening of the EIT resonance and, for example, negatively affect the memory performance.

In principle, it is possible to avoid the control absorption completely by arranging the frequencies of the stokes field and the Raman pump field to form a two-photon resonance between the $5S_{1/2},/F=2$ and $5D_{3/2}$ levels, as shown in Fig.~\ref{fig:eit_levels}(b). However, due to larger detuning from the intermediate excited level, this configuration leads to weaker Raman absorption. We were not able to observe more than $15\%$ stokes absorption even at maximum available pump power ($\approx 250$~mW).

Thus, we had to use a two-isotope configuration shown in Fig.~\ref{fig:eit_levels}(c), using $5S_{1/2},/F=2 \rightarrow 5S_{1/2},/F^\prime=2 \rightarrow 5D_{3/2}$ levels in ${}^{87}$Rb for stokes absorption.  This transition corresponds to the Raman pump wavelength of $762.0995$~nm. The sample stokes field absorption is shown in Fig.~\ref{fig:eit_levels}(b) for the Raman pump power $220$~mW. It is easy to observe multiple absorption resonances, due to the hyperfine structure of the $5D_{3/2}$ excited state, unresolved under the Doppler broadening. Typically, we tuned to the strongest Raman absorption peak.

\begin{figure}[H]
\begin{center}
	\includegraphics[width=1\columnwidth]{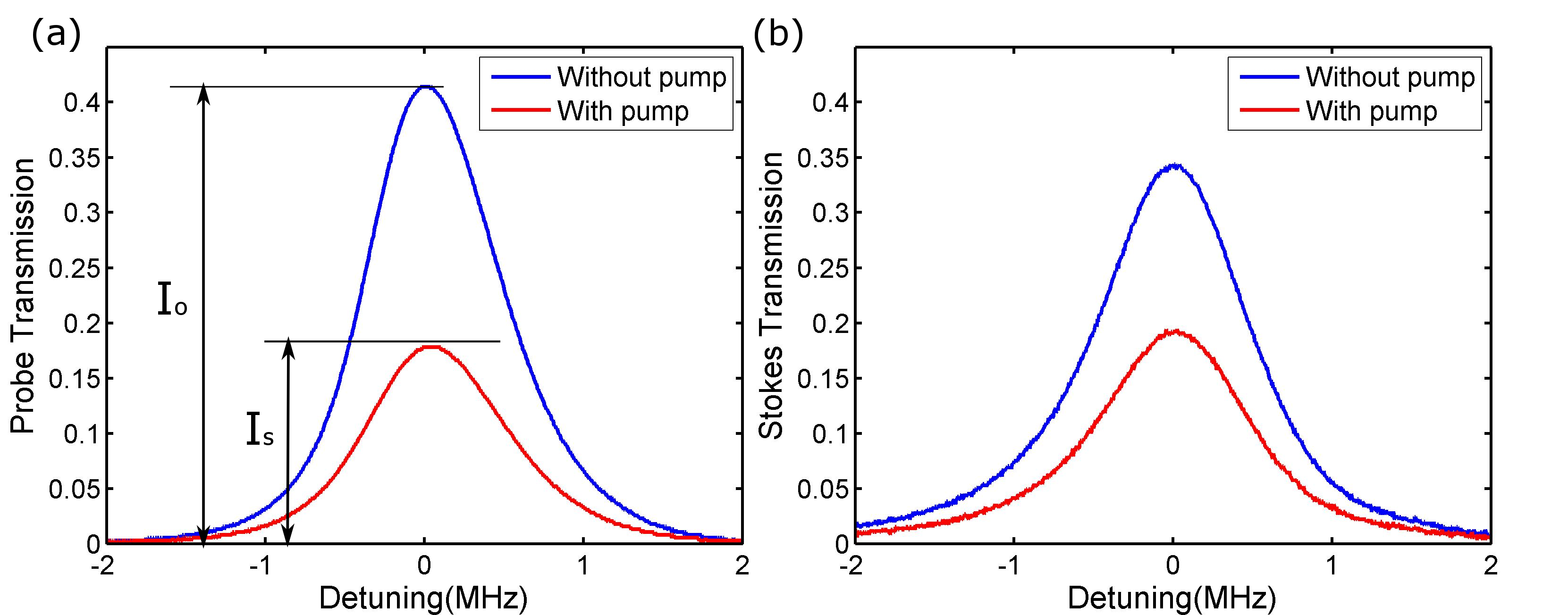}
	\caption{Transmission of \emph{(a)} probe field and \emph{(b)} stokes field as functions of two-photon frequency difference between the control and probe field with and without Raman pump-induced stokes absorption. All curves are normalized to the input probe field power. Raman pump power for both graphs is $220$~mW. Here $I_0$ and $I_S$ are the heights of the probe transmission peak without and with Raman pump, correspondingly. 
	 }
	\label{fig:Abs_sample_EIT}
\end{center}
\end{figure}

To study the effect of the stokes field absorption on the EIT/FWM, we recorded the variation in the output probe and stokes field when Raman pump laser was introduced.  The example of its effect on the output probe field is shown in Fig.~\ref{fig:Abs_sample_EIT}. As expected, we see the reduction of the probe transmission peak when the stokes field is absorbed (we have verified that Raman pump field does not directly affect probe propagation). Note that the stokes absorption did not affect the width of the transmission resonances, indicating that the observed peak reduction was not due to the deterioration of  the ground-state coherence.

\begin{figure}[H]
	\begin{center}
		\includegraphics[width=1\columnwidth]{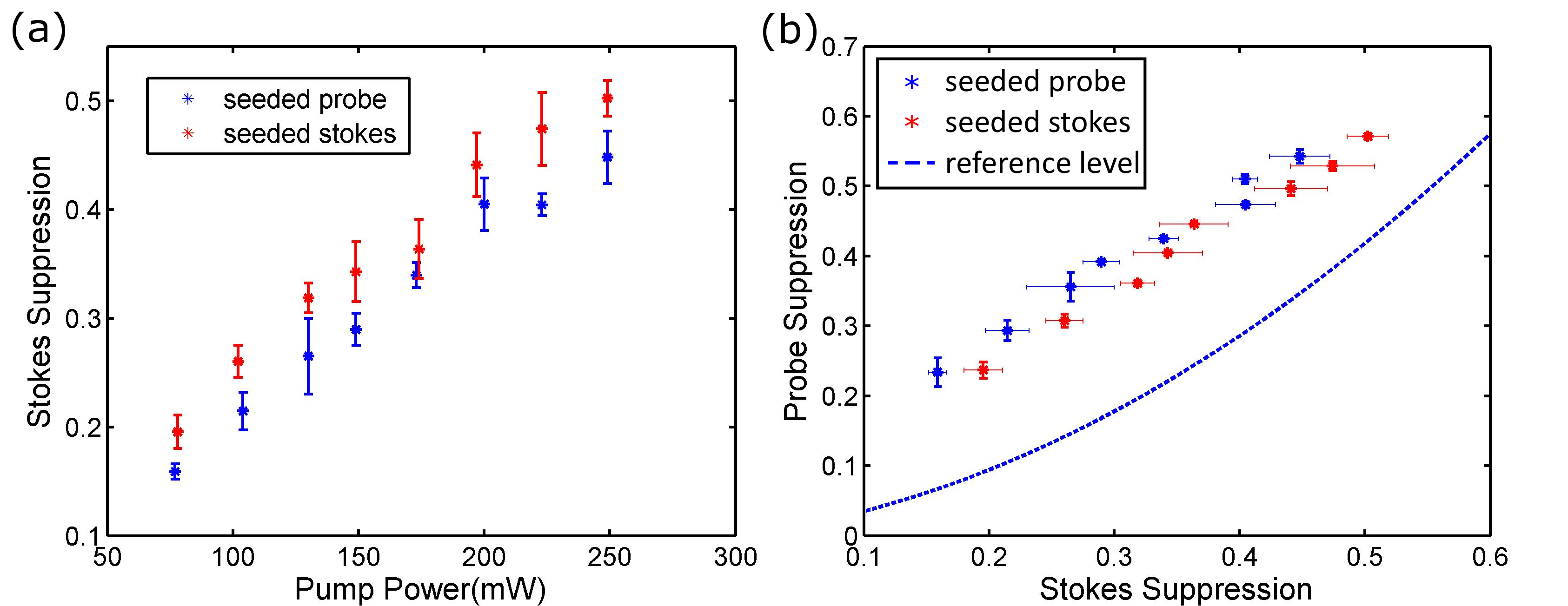}
		\caption{\emph{(a)} Suppression factor for stokes output field as a function of Raman pump power for EIT configuration. \emph{(b)} Probe field suppression as a function of the stokes field suppression. The data shown in red correspond to seeded input probe field (no input stokes), while the blue data points correspond to the seeded input stokes fields (no input probe). Each data point corresponds to the average of five independent measurements, with the error bar equal one standard deviation. Dashed blue line corresponds to the reference reduction factor for the output probe field, if the input stokes field was attenuated \emph{before} entering the vapor cell. }
		\label{fig:eit_graphs}
	\end{center}
\end{figure}

 To quantitatively characterize the effect of the Raman pump field, we calculated the suppression factor, defined as $1-\frac{I_S}{I_0}$, where $I_S$ and $I_0$ are the heights of the probe transmission peaks values with and without Raman pump, correspondingly (see the exact definitions in Fig.~\ref{fig:Abs_sample_EIT}(a).
Ideally, we would like to achieve unity suppression for the stokes field (i.e., no output stokes even at the two-photon resonance); since we ran the experiment at rather high Rb density, we also expect the probe suppression factor to approach one, as we expect only weak transmission due to exclusively EIT process.

The measured suppression factors for probe and stokes field as functions of Raman pump power are shown in Fig.~\ref{fig:eit_graphs}. We see that stokes absorption increases (roughly linear) with the Raman pump power, and the probe attenuation was linearly proportional to the stokes attenuation. In addition to a standard EIT arrangements, when a strong control field and a weak but nonzero probe field were injected into the Rb cell (no input stokes field), we also tested a configuration in which an optical field on the stokes frequency was injected, without any input probe field. In the letter case only FWM contributed toward the probe field observed after the interaction with the atoms. For this configuration we observed qualitatively similar behavior, although for the same pump power the stokes absorption was somewhat stronger. The limited available Raman pump power ($<250$~mW) did not allow us to reach the stokes field absorption beyond $50\%$, and corresponding probe suppression better than $60\%$. However, if we extrapolate the absorption data to the region of the higher pump powers, we can extrapolate that at a Raman pump power of $\approx 380$~mW we should be able to achieve optical depth $>1$ for the stokes field.

In addition to inducing stokes absorption via Raman resonance inside the vapor cell, we also measured the reduction in the output probe field as function of the input seeded stokes attenuation. These measurements are shown as a reference in Fig.~\ref{fig:eit_graphs}(b). It is easy to see that smaller probe suppression occurred in this case. The observed results can be explained by pointing out that in case of the seeded stokes its absorption can have two effects on probe. First, since the probe field is generated, its amplitude is proportional to the seeded stokes field, so weaker stokes is expected to produce less probe. This type of probe suppression should occur independently if the stokes field is attenuated before entering the cell or inside the interaction region. At the same time, additional stokes absorption can reduce the efficiency of the four-wave mixing, resulting in additional probe gain suppression, which maybe responsible for observed stronger suppression factor values in case of Raman absorption.

\section{Off-resonant Raman case}

\begin{figure}[H]
	\begin{center}
		\includegraphics[width=1.0\columnwidth]{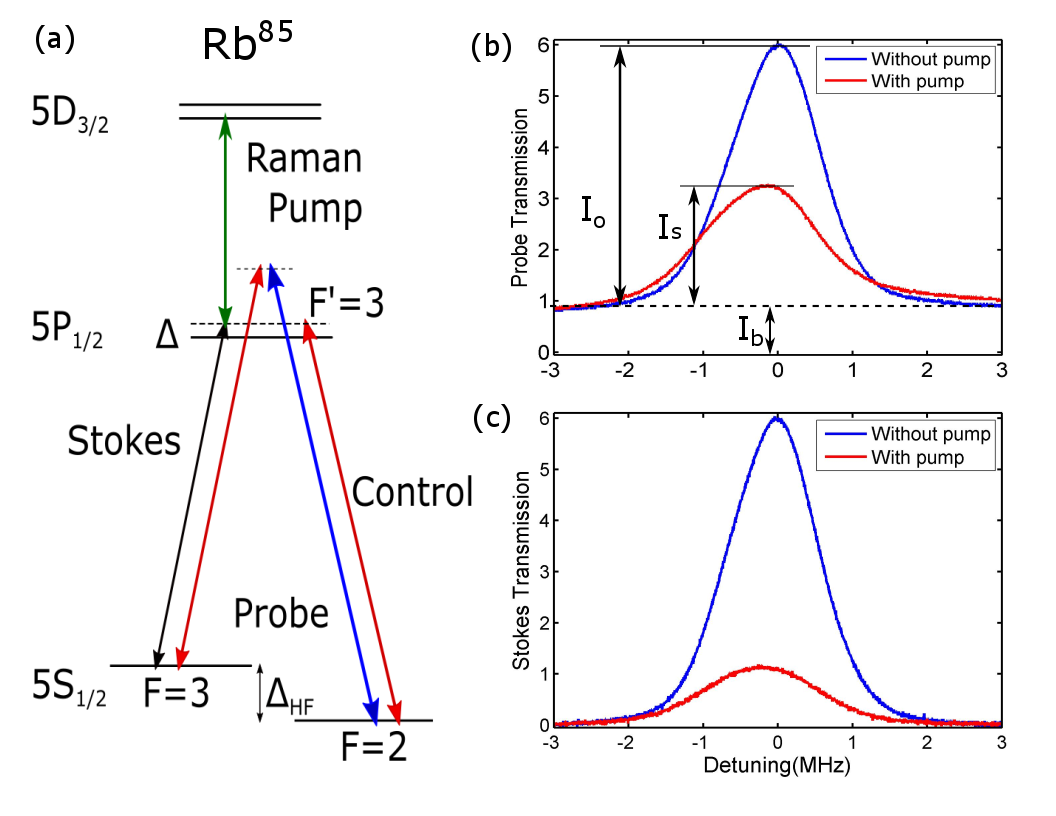}
		\caption{\emph{(a)} Level configuration used for the FWM suppression in the Raman configuration. The control field one-photon detuning from the $F=3\rightarrow F'=3$ transition is $\Delta = 200$~MHz, and the Raman pump field wavelength is $762.1038$~nm.
	\emph{(b, c)} Examples of the two-photon resonances for the probe and stokes fields with and without Raman pump of $80$~mW, correspondingly. All curves are normalized to the input probe field power. Here again $I_0$ and $I_S$ are the heights of the probe transmission peak without and with Raman pump, correspondingly, and $I_b$ is the background level, corresponding to the probe transmission away from the two-photon resonance.}
		\label{fig:raman_levels}
	\end{center}
\end{figure}

Another configuration identified as a promising candidate for the quantum memory applications is coherent Raman absorption of a probe field in a far-detuned $\Lambda$ system~\cite{nunnPRA2007,gorshkovPRA4,ReimPRL11}. This scheme also suffers from the effects of four-wave mixing noise~\cite{MichelbergerNJP15,quantmemoryreviewJMO2016}. While our limited laser power and cw regime of laser operation did not allow us to test the exact range of parameters used in the Raman memory experiments, we replicated their experimental arrangements as closely as possible. In particular, we have detuned both control and probe fields away from the atomic resonances, adding a one-photon detuning on the order of the hyperfine splitting between the ground state levels, as shown in Fig.~\ref{fig:raman_levels}(a). In this configuration the control field frequency approached the $F=2 \rightarrow F'=3$ transition, and the stokes field was generated near the $F=3 \rightarrow F'=3$ transition. Unlike in the EIT case, discussed above, there were very little ($>10\%$) resonant absorption for the probe field. At the same time, the stokes field experienced a rather stong resonant absorption due to the proximity to the optical resonance. If only seeded stokes field was interacting with atoms, it was nearly is completely absorbed. However, due to large FWM gain a significant generation (or enhancement) of the stokes field was observed after the Rb cell near the two-photon resonant conditions. 

\begin{figure}[H]
  \begin{center}
		\includegraphics[width=.6\columnwidth]{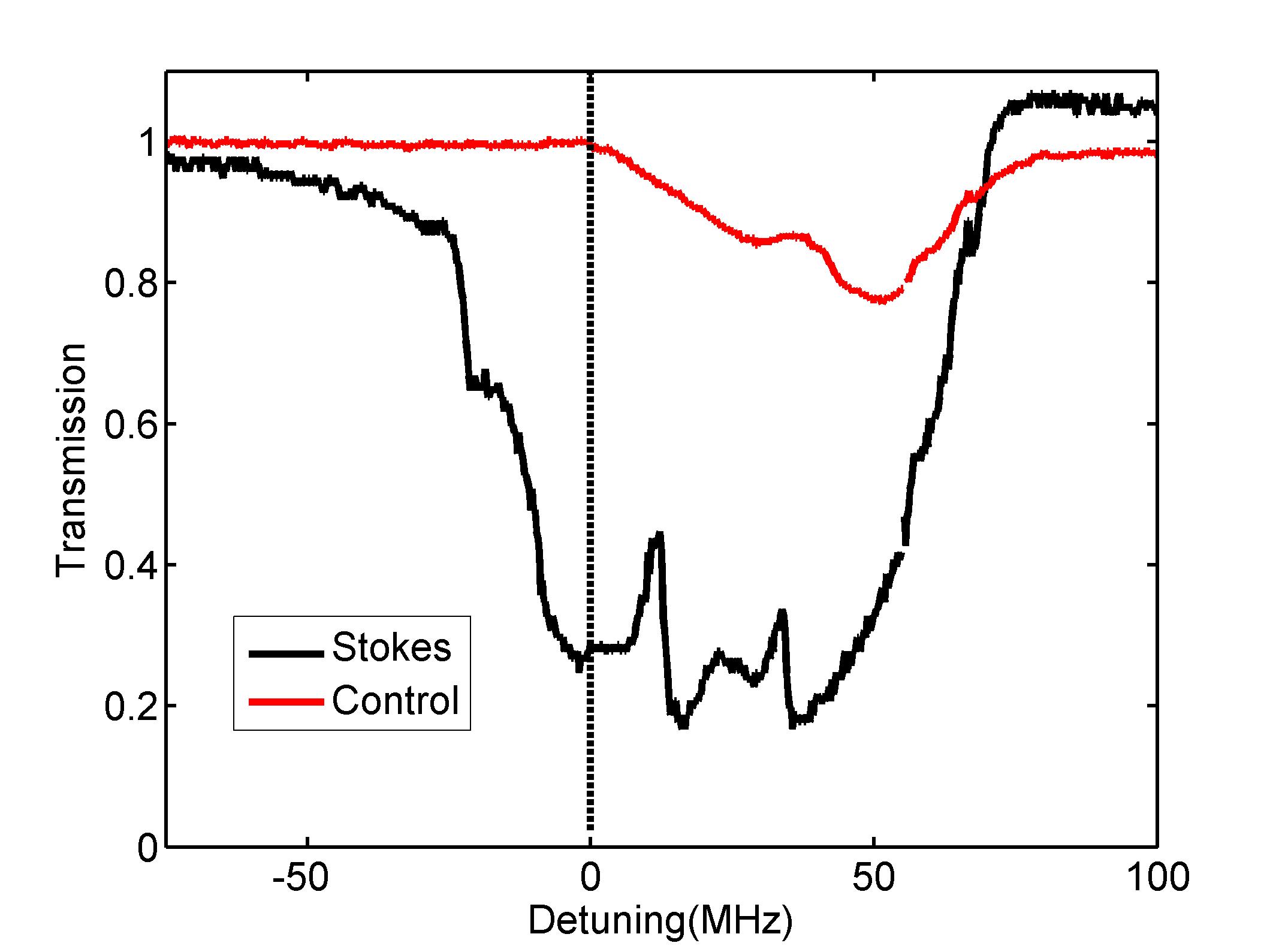}
		\caption{Measured transmission for the stokes and control optical fields as the Raman pump frequency is scanned across the absorption resonances. Raman pump power is $80$~mW. Vertical dashed line indicates the optimal operational frequency.}
		\label{fig:control_abs}
	\end{center}
\end{figure}

This configuration also allowed us to take advantage of the hyperfine structure of the $5D_{3/2}$ state to fine-tune the Raman pump frequency to absorb just the stokes field, with minimal control absorption. Example absorption profiles for the control and generated stokes field under the two-photon resonance conditions are shown in Fig.~\ref{fig:control_abs}. It is easy to see that the stokes absorption resonance, corresponding to the lowest pump frequency provides near-maximum stokes signal reduction, while keeping the control absorption less than $5\%$.

In this regime we were able to achieve much more significant levels of FWM suppression: nearly $95\%$ attenuation for the output stokes field at the highest Raman pump power. This more efficient absorption was likely due to the closer proximity of the stokes frequency to that of the optical resonance. As a result, much higher suppression was observed for lower powers, reaching the absorption saturation near half of the maximum power level. In this configuration,  we saw roughly the same amount of suppression for the stokes field when either probe or stokes were seeded. The suppression factor for the seeded probe was somewhat lower ($\approx 60\%$) compare to the seeded stokes case ($>75\%$). This reduction was somewhat expected: in case of the seeded stokes field, any output probe field is generated via the four-wave mixing process, and in case of perfect FWM suppression should vanish completely, resulting in the unity suppression factor. However, for seeded probe we expect to see a non-vanishing two-photon EIT resonance even if FWM completely eliminated, leaving the final suppression factor value below one. We also note that in this Raman regime the attenuation of the seeded stokes field either before or inside the cell gave similar generated probe suppression.
\begin{figure}[H]
	\begin{center}
		\includegraphics[width=1\columnwidth]{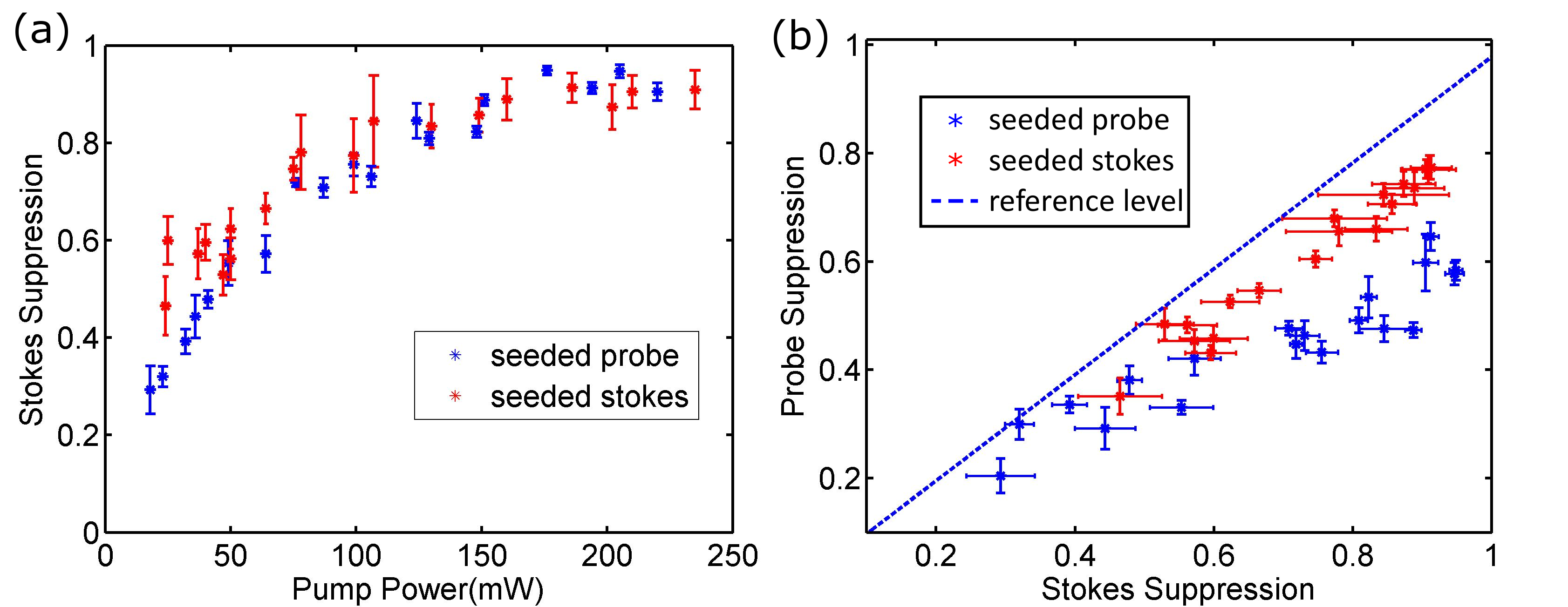}
		\caption{\emph{(a)} Suppression factor for stokes output field as a function of Raman pump power for far-detuned Raman configuration. \emph{(b)} Probe field suppression as a function of the stokes field suppression. The data shown in red correspond to seeded input probe field (no input stokes), while the blue data points correspond to the seeded input stokes fields (no input probe). Each data point corresponds to the average of five independent measurements, with the error bar equal one standard deviation. Dashed blue line corresponds to the reference reduction factor for the output probe field, if the input stokes field was attenuated \emph{before} entering the vapor cell. 
		}
		\label{fig: raman_graphs}
	\end{center}
\end{figure}

\section{Conclusion}

We demonstrated the possibility to use a ladder two-photon Raman absorption resonance to suppress four-wave mixing amplification of the probe field in a double-$\Lambda$ system under near-resonant EIT or far-detuned Raman conditions, the two interaction systems often considered for quantum memory experiments. We identified several configurations in which a strong optical field tuned in the vicinity of $5P_{1/2} \rightarrow 5D_{3/2}$ optical transition ($762$~nm) can produce narrow absorption resonances for the stokes optical field, generated in the four-wave mixing process. We showed substantial reduction in the output probe field when such resonances are introduced. Maximum four-wave mixing suppression in the EIT  configuration, based on ${}^{85}$Rb atoms, was approximately $40\%$ using the Raman resonance in ${}^{87}$Rb atoms. This value was limited by the available laser power. Same-isotope configurations were found as well, but either resulted in additional control field absorption, or required significantly stronger Raman pump field. In case of the far-detuned Raman double-$\Lambda$ system we achieved four-wave mixing suppression up to $85\%$ in the same ${}^{85}$Rb isotope, thanks to the stronger achievable stokes absorption.

\section{Aknowledgenets}

The authors thank  J. Nunn and A.~M. Akulshin for useful discussions.  This research was supported by AFOSR grant FA9550-13-1-0098.


\end{document}